\documentclass[10pt, conference, compsocconf]{IEEEtran}
\usepackage{cite}
\usepackage{amsmath,amssymb,amsfonts}
\usepackage{algorithmic}
\usepackage{graphicx}
\usepackage{textcomp}
\usepackage{xcolor}
\usepackage{url}
\usepackage{multirow}
\usepackage{booktabs}
\usepackage{mathtools}
\usepackage{listings}
\newcommand{\tabincell}[2]{\begin{tabular}{@{}#1@{}}#2\end{tabular}}

\begin{document}
\title{
ScaleHLS: A New Scalable High-Level Synthesis Framework on Multi-Level Intermediate Representation
\vspace{-10pt}
}


\author{
    \IEEEauthorblockN{
        Hanchen Ye$^1$, Cong Hao$^2$, Jianyi Cheng$^3$, Hyunmin Jeong$^1$, Jack Huang$^1$, Stephen Neuendorffer$^4$, Deming Chen$^1$
    }
    \IEEEauthorblockA{
        $^1$University of Illinois at Urbana-Champaign, $^2$Georgia Institute of Technology, $^3$Imperial College London, $^4$Xilinx Inc. \\
        \textit{hanchen8@illinois.edu, callie.hao@gatech.edu, jianyi.cheng17@imperial.ac.uk, hyunmin2@illinois.edu,} \\
        \textit{jackh4@illinois.edu, stephenn@xilinx.com, dchen@illinois.edu}
    }
    \vspace{-10pt}
}

\maketitle

\begin{abstract}
High-level synthesis (HLS) has been widely adopted as it significantly improves the hardware design productivity and enables efficient design space exploration (DSE). Existing HLS tools are built using compiler infrastructures largely based on a single-level abstraction, such as LLVM. However, as HLS designs typically come with intrinsic structural or functional hierarchies, different HLS optimization problems are often better solved with different levels of abstractions. This paper proposes \emph{ScaleHLS}\footnote{ScaleHLS is open-sourced at \url{https://github.com/hanchenye/scalehls}}, a new scalable and customizable HLS framework, on top of a multi-level compiler infrastructure called MLIR. ScaleHLS represents HLS designs at multiple representation levels and provides an HLS-dedicated analysis and transform library to solve the optimization problems at the suitable levels. Using this library, we provide a DSE engine to generate optimized HLS designs automatically. In addition, we develop an HLS C front-end and a C/C++ emission back-end to translate HLS designs into/from MLIR for enabling an end-to-end compilation flow. Experimental results show that, comparing to the baseline designs without manual directives insertion and code-rewriting, that are only optimized by Xilinx Vivado HLS, ScaleHLS improves the performances with amazing quality-of-results -- up to 768.1$\times$ better on computation kernel level programs and up to 3825.0$\times$ better on neural network models.
\end{abstract}


% \fi
% \IEEEpeerreviewmaketitle

\section{Introduction}
% Background and challenges of HLS.
High-level synthesis (HLS) automatically translates high-level languages into dedicated hardware accelerators, thereby removing the reliance of the cumbersome and potentially error-prone hardware design practices that use dedicated hardware description languages~\cite{rupnow2011study, kastner2018hls}. In recent years, HLS has been widely used in many application developments, such as neural networks~\cite{chen2019cloud, zhang2019skynet}, IoT applications~\cite{chen2016platform, zhang2015high, zhang2017machine}, and video processing~\cite{liu2016high}.  Existing algorithmic HLS tools typically focus on extracting parallelism from algorithmic descriptions and compiling the result into a parallel hardware execution model~\cite{papakonstantinou2009fcuda,papakonstantinou2011multilevel}. Thus, HLS tools would enable a designer to implement different algorithmic choices quickly, identify high-level area-performance tradeoffs, and avoid premature optimizations~\cite{cong2011hls}. While some of these alternatives can be explored automatically, it is also true that large-scale designs often make it very challenging to comprehensively explore the resulting large design space and produce high-quality design solutions~\cite{schafer2019high}. As a result, existing HLS tools often provide user-specified directives to control or guide the HLS process to generate different micro-architectures, which means the tools would rely on designers for writing 'good' code and setting 'good' directives to achieve good design quality~\cite{sohrabizadeh2020autodse}.

% Existing DSE efforts.
In recent years, we have witnessed many studies for investigating different design space exploration (DSE) methods of setting HLS directives~\cite{schafer2019high}. These efforts can be classified into two main types of methods: synthesis-based and model-based. Synthesis-based methods~\cite{schafer2009adaptive,cilardo2015interplay,ferretti2018lattice,sohrabizadeh2020autodse} invoke downstream HLS tools to evaluate the quality of result (QoR), including the latency, throughput, and resource utilization, of discovered design points. Model-based methods~\cite{zuo2015polyhedral, zhong2016lin, zhao2017comba, zuo2017accurate, wang2017flexcl} instead extract necessary design information from static dataflow graphs or dynamic execution traces and pass such information to predefined analytical models for estimating the QoR without invoking HLS tools. Recently, machine learning methods are also investigated ~\cite{o2018hlspredict, dai2018fast, makrani2019pyramid, wu2021ironman} to extract unique features that cannot be easily characterized by analytical models and deduce estimations for more complicated designs. Once performance and resource utilization estimates can be determined, the DSE process can be regularized and solved through simulated annealing~\cite{schafer2009adaptive}, linear programming~\cite{zuo2017accurate}, or other dedicated heuristics~\cite{ferretti2018lattice,sohrabizadeh2020autodse}, etc. Apart from different DSE methods, some other studies~\cite{papakonstantinou2009fcuda, shagrithaya2013enabling, lee2016openacc} leverage parallel-programming languages, such as CUDA~\cite{nickolls2008scalable}, as inputs to expose the parallelism of the accelerator designs and generate synthesizable C code with HLS directives inserted.

% Issues of existing efforts.
However, we find that existing research efforts and solutions face significant difficulty to handle large-scale HLS designs containing a large number of sub-modules and sophisticated inter-dependencies. The challenges mainly come from three aspects:

\textbf{Representation.} Existing works exploit C/C++ abstract syntax tree (AST)~\cite{lattner2008llvm}, traditional software compiler intermediate representation (IR)~\cite{lattner2004llvm}, or C/C++ source-level IR~\cite{dave2009cetus, lee2014openarc}, to represent and analyze HLS designs. These representations are originally designed for software compilation and only contain a single operation-level abstraction. However, HLS optimizations can often be carried out at or across different levels of abstraction for better results. For example, task/module level parallelization should be applied on high-level operators, such as convolution operators, rather than nested loops to avoid conservative assumption and sophisticated memory dependency analysis. Directly combining different levels of representation from different frameworks could cause significant fragmentation and cumbersome and inconsistent cross-level optimizations. We argue that we should have a systematic approach to represent HLS designs at multiple abstraction levels in order to honor the intrinsic hierarchies of HLS designs. This representation should act as the foundation of HLS optimization and address the various fragmentation and inconsistency issues that we are facing.

\textbf{Optimization.} Existing works leave many important HLS optimizations, such as task/module level resource-sharing and parallelization, hardware IP integration, and loop level analysis and transformation, to human designers done by manual code rewriting. Such an approach is not productive and scalable enough to deal with large HLS designs and may obstruct the comprehensive DSE. We argue that HLS optimizations should be fully automated and parameterized rather than relying on manual code rewriting. These optimizations should be carried out at multiple different abstraction levels automatically to reduce the complexity of program analysis and make the compilation flow more scalable to large HLS designs.

\textbf{Exploration.} In the domain of compiler development, the parameters of each optimization technique are typically determined by a \emph{cost model} indicating the 'benefit' of the combination of such parameters. However, in HLS designs, because the effects of different HLS optimizations correlate (and sometimes in conflict) with one another, we cannot calculate the 'benefit' of one optimization in isolation of the other optimizations. In order to solve this problem, a global DSE engine is desired to take all HLS optimizations across different levels of abstraction into consideration and explore the large design space effectively.

In this paper, we propose a new tool, named as \emph{ScaleHLS}, to tackle the challenges present in the representation, optimization, and exploration of HLS designs. ScaleHLS represents HLS designs with a multi-level IR for the first time, solves HLS optimization problems at the right levels of abstraction, and automates such optimizations through a new end-to-end flow. ScaleHLS can optimize large HLS designs and still deliver high QoR for FPGA hardware implementation. 
We summarize the main contributions of our work as follows.
\begin{itemize}
    \item To the best of our knowledge, ScaleHLS is the first end-to-end automated HLS compilation flow built on multiple levels of design abstraction naturally honoring intrinsic structural or functional hierarchies of large-scale designs.
    \item ScaleHLS proposes a hierarchical and scalable HLS representation and optimization methodology, which optimizes HLS designs at graph, loop, and directive levels holistically, to handle the complexity of the increasing HLS design space.
    \item ScaleHLS provides an HLS-dedicated transform and analysis library, which turns a set of HLS optimization techniques from manual code rewriting to callable and tunable interfaces, saves significant amount of human effort and establishes the foundation of automated DSE.
    \item ScaleHLS contains a novel automated DSE engine to search for the Pareto frontier of the latency-area tradeoff space. A QoR estimator is also developed to evaluate design points discovered by the DSE engine rapidly.
    \item ScaleHLS expands the MLIR framework by providing an HLS C front-end and a synthesizable HLS C/C++ emission back-end for bridging the gap between the multi-level IR and C-based HLS designs, thus enabling an end-to-end HLS compilation flow.
\end{itemize}

% Paper structure.
The remaining of this paper is organized as follows. Section \ref{sec:background} introduces the background. In Section \ref{sec:framework}, we provide an overview of the ScaleHLS framework. In Sections \ref{sec:representation} and \ref{sec:optimization}, we introduce the details of the multi-level representation and optimization for HLS designs, respectively. In Section \ref{sec:integration}, we present the front-end and back-end integration of ScaleHLS. In Sections \ref{sec:results} and \ref{sec:conclusion}, we provide the evaluation results and conclude this paper, respectively.

\begin{figure}
    \centering
    \includegraphics[width=\linewidth]{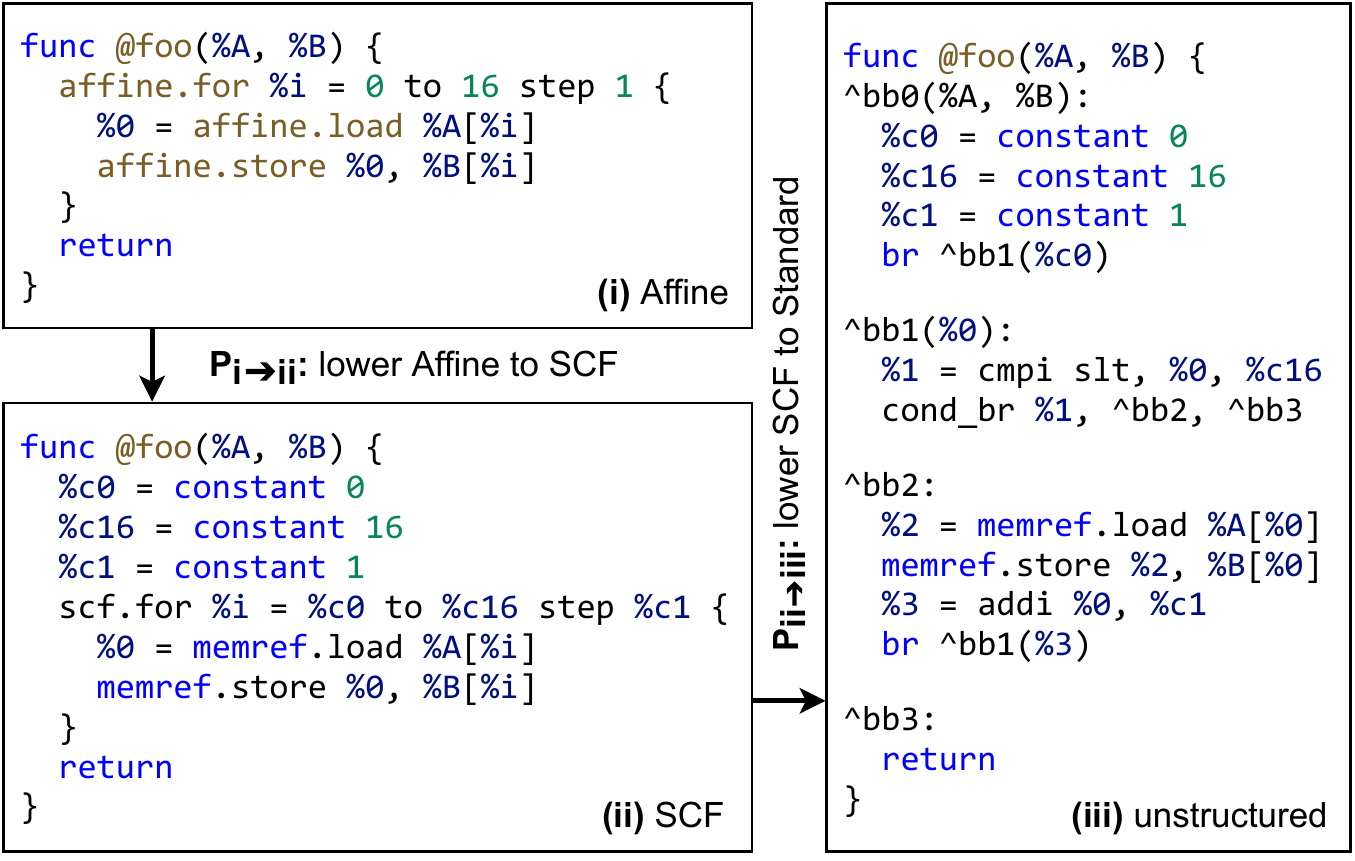}
    \vspace{-10pt}
    \caption{An IR example, where \texttt{affine} and \texttt{scf} dialect represents structured control flow. \texttt{affine} dialect can be lowered to \texttt{scf} and then lowered to unstructured IR. All types are omitted for simplicity.}
    \label{fig:mlir}
    \vspace{-5pt}
\end{figure}

\section{Background}
\label{sec:background}

\subsection{MLIR Framework}
ScaleHLS is built on top of MLIR~\cite{lattner2020mlir, mlirgithub}, a compilation framework supporting multiple levels of functional and representational hierarchy. In the remainder of this paper, we use \emph{MLIR} to refer to the MLIR compilation framework and \emph{IR} for the intermediate representation of programs in MLIR. MLIR includes a single static assignment (SSA) style IR~\cite{cytron1991efficiently} where an \emph{Operation} is the minimal unit of code. Each operation accepts a set of typed \emph{Operand}s and produces a set of typed \emph{Result}s. Connections between the results of one operation and the operands of another operation describe the SSA-style flow of data. For instance, \texttt{\%3 = addi \%0, \%c1} in Fig. \ref{fig:mlir}(iii) is an operation with operands \texttt{\%0} and \texttt{\%c1} and result \texttt{\%3}. Each operation can also be parameterized by a set of \emph{Attribute}s indicating important characteristics of the operation. Unlike operands, which typically model values produced by other operations when a program is executed, attributes have values that are known and fixed at compile time. A sequential list of operations without control flow is defined as a \emph{Block} and a control flow graph (CFG) of blocks is organized into a \emph{Region} in MLIR. Regions are, in turn, contained by operations, enabling the description of arbitrary design hierarchy. In MLIR, \emph{Function} is defined as a built-in callable operation always owning one region. For instance, function \texttt{@foo} in Fig. \ref{fig:mlir}(iii) owns one region containing four blocks, \texttt{bb0} to \texttt{bb3}.

A \emph{Dialect} in MLIR defines a namespace for a group of related operations, attributes, and types. MLIR not only provides multiple built-in dialects to represent common functionalities, but also features an open infrastructure allowing to define new dialects at different abstraction levels. \emph{Pass} is a key component of compiler which traverses the IR for the purpose of optimization or analysis. Similar to LLVM~\cite{lattner2004llvm}, users can design \emph{Transform} and \emph{Analysis} passes in MLIR to perform the IR transformation and analysis, respectively. However, in the context of MLIR, \emph{Transform} typically refers to the transformation within a dialect. The transformation between different dialects is typically referred as \emph{Conversion}, while the transformation between MLIR and external representation is referred as \emph{Translation}. \emph{Lowering} is a terminology referring to the process of lowering the abstraction level of IR.

\subsection{Relevant MLIR Dialects}
\label{subsec:mlir_dialects}
Many dialects in MLIR are immediately applicable for representing nested loop programs commonly used in HLS. The \texttt{affine} dialect provides a powerful abstraction for affine operations in order to make dependence analysis and loop transformations efficient and reliable. The \texttt{affine} dialect defines \emph{Affine Map} as a mathematical function that transforms a list of affine values into a list of results. Affine operations (e.g., \texttt{affine.for} and \texttt{if}) must take affine values as input operands, therefore the loop bounds of \texttt{affine.for} operation and conditions of \texttt{affine.if} operation must be the expression of affine values. The \texttt{scf} (structured control flow) dialect defines control flow operations (e.g., \texttt{scf.for} and \texttt{if}) whose loop bounds or conditions can be any SSA values. Therefore, \texttt{scf} operations are not constrained by the affine requirements and can represent a wider range of programs. MLIR also provides several fundamental built-in dialects to represent basic arithmetic operations (e.g., \texttt{addf}) and unstructured control flow operations (e.g., \texttt{br} and \texttt{cond\_br}). Taking Fig. \ref{fig:mlir} as example, the structured control flows in Fig. \ref{fig:mlir}(i) and (ii) represented with \texttt{affine} and \texttt{scf} operations are flattened to the unstructured \texttt{br} and \texttt{cond\_br} operations in Fig. \ref{fig:mlir}(iii).

\subsection{Relevant MLIR Front-ends}
ScaleHLS uses existing third-party front-ends, Torch-MLIR~\cite{torchmlirgithub} and ONNX-MLIR~\cite{le2020compiling}, to parse PyTorch~\cite{paszke2019pytorch} and ONNX~\cite{onnxgithub} models, respectively. Torch-MLIR first translates PyTorch models into \texttt{torch} dialect, then lowers the IR to \texttt{affine} dialect as the end of compilation. ONNX-MLIR defines a subset of ONNX operations in an \texttt{onnx} dialect for translating ONNX models into MLIR. The \texttt{onnx} operations are then lowered to \texttt{krnl} (kernel) dialect and finally lowered to \texttt{affine} dialect by the ONNX-MLIR compilation flow.

% ScaleHLS also includes its own C frontend for MLIR. The existing C front-end Polygeist~\cite{moses2021polygeist} requires the users to manually identify the affine region in C using \texttt{scop} pragmas, while our approach can take arbitrary C code and automatically identifies the affine region in MLIR. Also, most HLS applications contain partially affine loops. Their approach is an all-or-nothing process, which means that a non-affine statement in a given region can cause failure in translating the whole region to the \texttt{affine} dialect, while our approach supports more precise granularity and translates other affine statements in this region into \texttt{affine} operations.

\begin{figure}
    \centering
    \includegraphics[width=\linewidth]{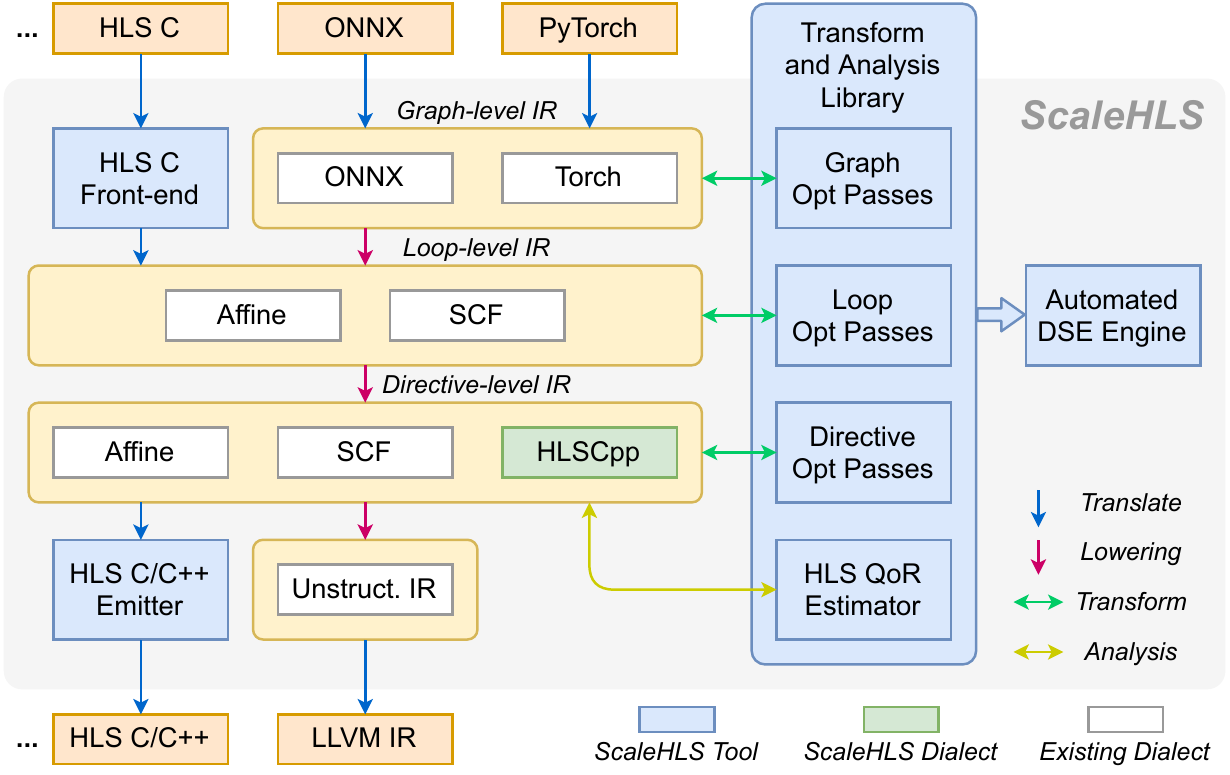}
    \vspace{-10pt}
    \caption{ScaleHLS framework.}
    \label{fig:framework}
    \vspace{-5pt}
\end{figure}

\section{ScaleHLS Framework Overview}
\label{sec:framework}
ScaleHLS compiles programs described in HLS C code or programming frameworks, such as ONNX and PyTorch, to optimized and synthesizable HLS C/C++ designs. Fig. \ref{fig:framework} shows the architecture of ScaleHLS. In this section, we organize the main components into four categories (representation, optimization, exploration, and integration) and introduce them one by one.

\subsection{Representation}
\textbf{Graph-level IR} (Section \ref{subsec:graph_ir}) exploits existing third-party \texttt{onnx}~\cite{le2020compiling} dialect to represent computation graphs constructed with tensor operations. \textbf{Loop-level IR} (Section \ref{subsec:loop_ir}) exploits MLIR built-in \texttt{affine} and \texttt{scf} dialects to represent the loop structures and leverage the powerful loop transformation and analysis infrastructures of MLIR. \textbf{Directive-level IR} (Section \ref{subsec:hls_directive}) is enabled by our customized \texttt{hlscpp} dialect to represent the HLS-specific structures and program directives, which provides the capability of conducting directive optimizations and supports the emission of synthesizable C/C++ code.

\subsection{Optimization}
On each level of IR, we have designed a set of \textbf{Optimization passes} (Section \ref{subsec:graph_transform} to \ref{subsec:simplification}) to improve the HLS design quality automatically. The hierarchical IR allows the passes to be applied at the most suitable abstraction level, thereby minimizing the processing complexity and improving the scalability. In order to efficiently explore the large design spaces brought by large HLS designs, we propose a fast \textbf{HLS QoR estimator} (Section \ref{subsec:qor_estimation}) based on analytical models, which can estimate the latency and resource utilization of programs on top of the structured directive-level IR.

\subsection{Exploration}
The interfaces of the QoR estimator and transform passes of every abstraction level are packaged into an \textbf{HLS transform and analysis library} (Section \ref{sec:optimization}). All the interfaces in the library are highly parameterized and can be tuned by DSE engines. This library turns the HLS optimization techniques from manual code rewriting to callable and tunable interfaces at different abstraction levels. Leveraging the HLS transform and analysis library, we have designed an \textbf{Automated DSE engine} (Section \ref{subsec:dse_algorithm}) to search for the Pareto frontier of the multi-dimensional design space, where each dimension corresponds to a tunable parameter of a transform pass. The DSE engine is extensible to support different optimization algorithms in the future.

\subsection{Integration}
We have implemented an \textbf{HLS C front-end} (Section \ref{subsec:front_end}) based on Clang that directly translates input C programs into the \texttt{scf} dialect. An \texttt{scf} to \texttt{affine} raising pass identifies affine regions and converts \texttt{scf} operations to their corresponding \texttt{affine} operations automatically, which enables subsequent affine transformations and analyses. In the end of compilation, the structured directive-level IR is translated into synthesizable C++ code by an \textbf{HLS C/C++ emitter} (Section \ref{subsec:emitter}). Meanwhile, LLVM IR~\cite{lattner2004llvm} can also be generated, enabling software simulation and direct interfacing with existing LLVM-compatible tools, such as Xilinx Vitis HLS \cite{vitishlsgithub}.

\section{ScaleHLS Representation}
\label{sec:representation}
ScaleHLS features an unique multi-level representation which allows the transform and analysis passes to be applied on multiple abstraction levels, thereby exploring more comprehensive design spaces and improving scalability. In this section, we introduce the graph, loop, and directive level IRs of ScaleHLS in detail.

\subsection{Graph-level IR}
\label{subsec:graph_ir}
ScaleHLS exploits existing third-party \texttt{onnx} dialect from ONNX-MLIR~\cite{le2020compiling} to represent and optimize graph-level IR. The assembly form of an \texttt{onnx.Conv} operation is (attributes and non-tensor operands are omitted for simplicity):

\begin{footnotesize}
\begin{verbatim}
%output = "onnx.Conv"(%input, %weight) {...} :
  (tensor<1x3x34x34xf32>, tensor<64x3x3x3xf32>)
  -> tensor<1x64x32x32xf32>
\end{verbatim}
\end{footnotesize}
where the matrix operands and result are typed as tensors. Operations of this dialect consume and produce tensor-type values, which allows optimizing the IR at this level through simple define-use analysis. If these operations are lowered to loop-level and tensors are bufferized to memories, tensor data must be accessed through memory read and write operations, making optimization and transformation more cumbersome due to the need for sophisticated memory dependency analysis.  In contrast, many high level analyses and transformations, such as graph node merging, can be easily supported in a graph-level IR by manipulating tensor operations. The graph-level transformations implemented in ScaleHLS are discussed further in Section \ref{subsec:graph_transform}.

\subsection{Loop-level IR}
\label{subsec:loop_ir}
Once the graph-level optimizations are completed, the IR will be lowered to loop-level for further optimization. ScaleHLS exploits the MLIR built-in dialects, particularly \texttt{affine} and \texttt{scf}, to represent loop-level IR for reusing the powerful analysis and transform libraries provided by MLIR. The code block (ii) of Fig. \ref{fig:opt} shows the loop-level IR of an SYRK (symmetric rank-k update of a matrix) computation kernel~\cite{blackford2002updated} in MLIR where types and attributes of all operations are omitted for simplicity. Memory access and arithmetic operations are nested in \texttt{affine.for} operations, which explicitly represent the loop structures. Similarly, the code block (iii) of Fig. \ref{fig:opt} shows the structured representation of a conditionally executed MLIR block contained by an \texttt{affine.if} operation. Compared to the unstructured IR, the structured loop-level IR enables more flexible and efficient loop optimizations (e.g., loop tiling). Furthermore, the fast affine expression composition and the use of affine transformation theory allow ScaleHLS to perform efficient and comprehensive analyses and transformations on \texttt{affine} operations. The loop-level optimizations are discussed in detail in Section \ref{subsec:loop_transform}.

\subsection{HLS Directives}
\label{subsec:hls_directive}
HLS tools typically use program directives to guide the hardware generation and fine-tune the latency-area tradeoff. In this section, we introduce how ScaleHLS represents the function, loop, and array HLS directives shown in Tab. \ref{tab:directives}.

\begin{table}
    \centering
    \caption{Supported HLS directives.}
    \vspace{-5pt}
    \begin{tabular}{cccc}
        \toprule
         & \textbf{Function} & \textbf{Loop} & \textbf{Array} \\
        \midrule
        \textbf{Directives} & 
        \tabincell{c}{dataflow \\ pipeline \\ inline} & 
        \tabincell{c}{dataflow \\ pipeline \\ unroll \\ merge} & 
        \tabincell{c}{partition \\ resource \\ interface} \\
        \bottomrule
    \end{tabular}
    \vspace{-5pt}
    \label{tab:directives}
\end{table}

\subsubsection{Function Directives}
\label{subsec:func_directive}
ScaleHLS supports coarse-grained and fine-grained parallelism through applying directives. The \emph{dataflow} directive enables task parallelism by pipelining all sub-functions that appear in the target function. In the generated hardware, the top-module will be ready to accept a new frame of data once the first sub-module is done, which effectively improves the throughput of the top-module. The \emph{pipeline} directive enables operation parallelism by scheduling all operations in the target function into multiple pipelined stages that can be executed in parallel. For the \emph{pipeline} directive, ScaleHLS allows specifying the targeted initiation interval ($II$), which indicates that the pipeline can accept and process a new input every $II$ clock cycles, impacting the resource usage and performance of the generated pipeline. To represent and parse these directives in ScaleHLS, we customize a \texttt{struct} MLIR attribute named \texttt{FuncDirective} in \texttt{hlscpp} dialect. The customized attribute contains two Boolean parameters, \texttt{dataflow} and \texttt{pipeline}, and one integer parameter, \texttt{targetII}, which triggers the generation of appropriate directives compatible with downstream HLS tools, such as Xilinx Vivado HLS \cite{hls2020userguide}. In ScaleHLS, the function \emph{inline} directive is not explicitly represented with an MLIR attribute, but instead directly inlines the target function in the IR to ease the transformation and analysis.

\subsubsection{Loop Directives}
\label{subsec:loop_directive}
The throughput and latency of loop regions can also be optimized by applying the \emph{dataflow} and \emph{pipeline} directives, which largely share the same characteristics with the corresponding function directives. Note that ScaleHLS can automatically identify perfectly nested loops and flatten them into a single hierarchy, which helps to further improve the pipeline throughput and latency. Similar to function directives, ScaleHLS also exploits customized MLIR attributes to represent the loop \emph{dataflow} and \emph{pipeline} directives and the targeted $II$. A \texttt{LoopDirective} attribute is defined in \texttt{hlscpp} dialect and attached to the corresponding \texttt{affine.for} or \texttt{scf.for} operations when directives are applied.

The computation parallelism of loops can be improved by applying the loop \emph{unroll} directive with the cost of consuming more resources. The \emph{merge} directive is used to fuse adjacent loop nests to improve data locality and decrease the loop control overhead. ScaleHLS does not explicitly represent these two directives through MLIR attributes, but instead directly performs corresponding loop transformation on the target loops in the IR, which is semantically equivalent to applying the directives.

\begin{figure}
    \centering
    \includegraphics[width=\linewidth]{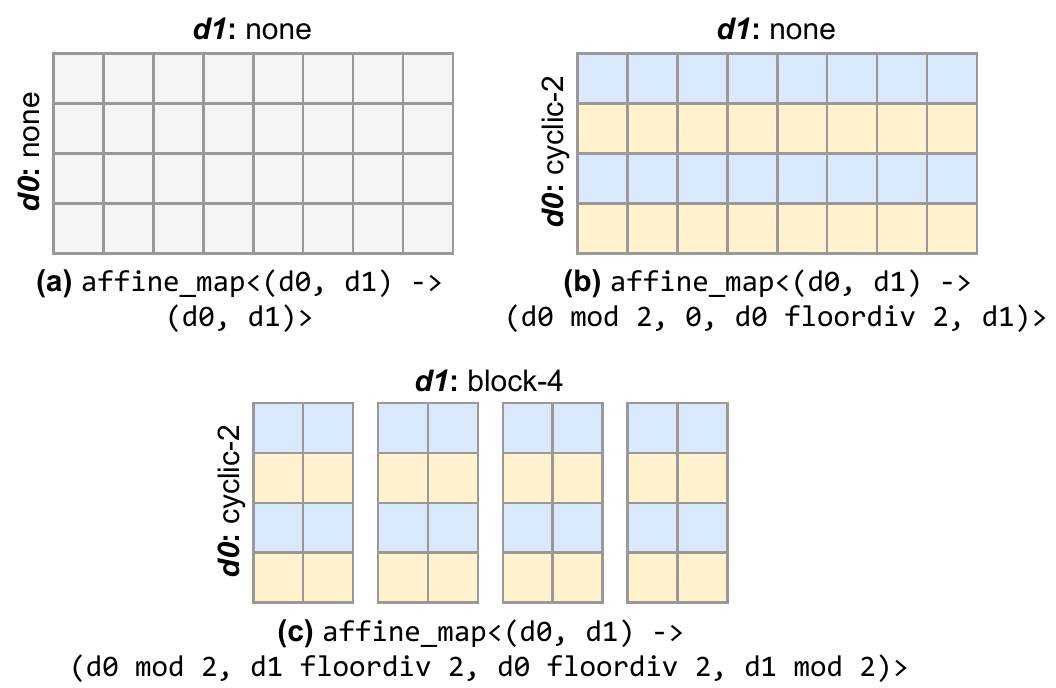}
    \vspace{-10pt}
    \caption{Affine-based array partition. $d\{n\}$ indicates the $n$-th dimension of the array. Partition fashions and factors: (a) without partition; (b) cyclic partitioned along the dimension-0 with a factor of 2; (c) block partitioned along dimension-1 with a factor of 4.}
    \label{fig:partition}
    \vspace{-5pt}
\end{figure}

\subsubsection{Array Partition}
Array partition is one of the most important HLS directives because HLS designs require enough on-chip memory bandwidth to comply with the computation parallelism. However, single on-chip memory block has limited read/write ports and hence needs to be partitioned into multiple physical blocks to enable massive simultaneous read and write. As MLIR attaches an affine map to each memory type for encoding the memory layout, ScaleHLS reuses the affine-based memory typing system of MLIR to flexibly represent the partition factor (the number of memory blocks after partition) and various partition fashions (e.g., \emph{cyclic} and \emph{block}). Fig. \ref{fig:partition} shows three examples including: (a) array without partition, (b) partitioned along the first dimension, and (c) partitioned along both two dimensions. The partition fashions and factors and the corresponding affine map are annotated to each example as well. As we introduced in Section \ref{subsec:mlir_dialects}, affine map is a transform function mapping a list of affine inputs to a list of results. To represent array partition in ScaleHLS, assuming an $N$-dimensional target array, the attached affine map has $N$ inputs and $2N$ results. While the inputs are the logical indices of the array, the first and last $N$ results are used to encode the expressions of partition indices and physical indices after array partition, respectively. Taking the affine map of Fig. \ref{fig:partition}(b) as an example, the partition index and physical index of $d0$ can be calculated as $d0~\%~2$ and $\lfloor d0~/~2 \rfloor$ when  dimension-0 is partitioned cyclically with a factor of two.

By encoding the partition information into memory types, ScaleHLS can flexibly support different partition fashions, and quickly infer the partition index and physical index through affine expression composition. This technique is used in the QoR estimator (Section \ref{subsec:qor_estimation}) and the \texttt{-array-partition} pass (Section \ref{subsec:array_partition}). Note that unsupported memory partition fashions by the downstream HLS tools are disallowed in the directive-level IR of ScaleHLS.

\begin{table}
    \centering
    \caption{ScaleHLS passes.}
    \vspace{-5pt}
    \begin{tabular}{c|l|l|l}
        \midrule
        & \multicolumn{1}{c|}{\textbf{Passes}} & \multicolumn{1}{c|}{\textbf{Target}} & \multicolumn{1}{c}{\textbf{Parameters}} \\
        \midrule
        \multirow{2}{*}{\textbf{Graph}}
        & \textbf{-legalize-dataflow} & function & insert-copy \\
        & \textbf{-split-function} & function & min-gran \\
        \midrule
        \multirow{6}{*}{\textbf{Loop}}
        & \textbf{-affine-loop-perfectization} & loop band & - \\
        & \textbf{-affine-loop-order-opt} & loop band & perm-map \\
        & \textbf{-remove-variable-bound} & loop band & - \\
        & -affine-loop-tile & loop band & tile-sizes \\
        & -affine-loop-unroll & loop & unroll-factor \\
        % & -affine-loop-fusion & loop & - \\
        \midrule
        \multirow{3}{*}{\textbf{Direct.}}
        & \textbf{-loop-pipelining} & loop & target-ii \\
        & \textbf{-func-pipelining} & function & target-ii \\
        & \textbf{-array-partition} & function & part-factors \\
        \midrule
        \multirow{4}{*}{\tabincell{c}{\textbf{Misc.}}}
        & \textbf{-simplify-affine-if} & function & - \\
        & \textbf{-affine-store-forward} & function & - \\
        & \textbf{-simplify-memref-access} & function & - \\
        & -canonicalize -cse & function & - \\
        \midrule
        \multicolumn{4}{l}{\tabincell{l}{Boldface ones are new passes provided by ScaleHLS, while others \\ are MLIR built-in passes.}}
    \end{tabular}
    \label{tab:passes}
    \vspace{-5pt}
\end{table}

\subsubsection{Array Resource and Interface} The HLS-based accelerators can use different kinds of memories, including on-chip memories (e.g., BRAM) and off-chip memories (e.g., DRAM). The resource directive is introduced for indicating what kind of memories should an array be allocated to. This is similar to the concept of memory space in the software, where BRAM and DRAM respond to cache and main memory of a common computer system. As MLIR also encodes the memory space into the memory type system, ScaleHLS reuses this for representing resource directive by mapping different kinds of memories into different memory spaces. Notably, ScaleHLS distinguishes single port, simple dual-port, and true dual-port on-chip memories to precisely control the resource utilization. Additionally, if an array is identified as a function argument or returned value, ScaleHLS will automatically determine the interface category (e.g., AXI~\cite{axi2017guide} or naive BRAM interface) of the array according to its memory space.

\section{ScaleHLS Optimization}
\label{sec:optimization}
On top of the hierarchical representation of ScaleHLS, we propose a multi-level HLS optimization methodology to address the challenges of optimizing large HLS designs. This methodology is implemented using a set of MLIR transformation passes, each operating on MLIR dialects at an appropriate abstraction level, either the graph, loop, or directive levels described above. All ScaleHLS transform passes are listed in Tab. \ref{tab:passes}, together with their transform targets (e.g., function) and the tunable parameters, where a \emph{Loop Band} refers to a continuous set of loops. These passes traverse the whole IR and operate on all suitable targets in the IR, making it difficult to apply different combinations of passes on different targets through the command line tool. To solve this problem, we also expose the functionality of each transform pass as a callable method, allowing precise control on where transforms are applied. These methods together with the QoR estimator are packaged into an HLS transform and analysis library, which opens the opportunity to perform comprehensive DSEs by applying different combinations of transforms on different targets in the IR and tuning their parameters.

In this section, we first introduce the graph, loop, and directive passes accordingly. Then, we introduce other transform passes provided by ScaleHLS for eliminating redundancies. Finally, we introduce the QoR estimator and the automated DSE algorithm. In addition, downstream HLS tools are observed unpredictable: changing parameters in the program that should improve performance can counter intuitively yield slower and larger designs~\cite{nigam2020predictable}. ScaleHLS deals with this problem with predictable transform passes and the integrated QoR estimator, which will be elaborated further in this section.

\subsection{Graph Transform Passes}
\label{subsec:graph_transform}
\subsubsection{Legalize Dataflow}
Downstream HLS tools often support dataflow pipelining with specific restrictions in coding style. In particular, for Vivado HLS each intermediate result must have only one producer and one consumer, bypass and feedback paths are not allowed, and conditional executions of sub-functions are not allowed~\cite{hls2020userguide}. Previously, users were required to manually legalize the target function by splitting the function body into multiple sub-functions and rewriting the code structure to eliminate the bypass, multi-producer, or multi-consumer data paths. This procedure is (1) error-prone and unpredictable since careless rewriting can easily result in incorrect functionality or undesired effect in terms of performance and resource utilization and (2) less effective since large HLS designs containing tens of sub-functions can take up to hours for human designers to reorganize and split. The drawbacks of such manual efforts obstruct the existing HLS tools to effectively explore different configurations of the dataflow pipelining. Previous work \cite{voitsechov2014single} proposes an automatic method to enable thread-level dataflow on GPGPUs, yet the dataflow issue in HLS designs has not been well-studied.

\begin{figure}
    \centering
    \includegraphics[width=\linewidth]{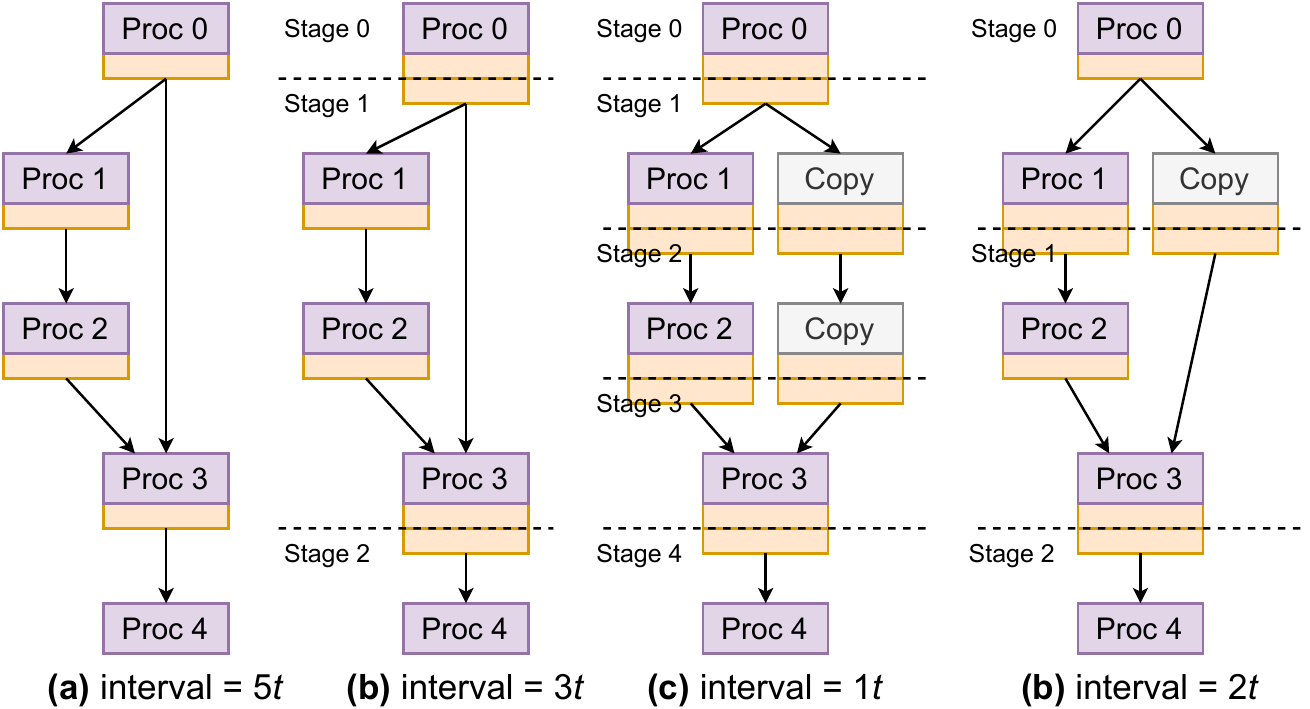}
    \vspace{-10pt}
    \caption{Graph-level dataflow optimization. (a) original dataflow; (b) legalized dataflow without copy nodes; (c) legalized dataflow with inserting copy nodes; (d) dataflow with a minimum granularity of 2.}
    \label{fig:dataflow}
    \vspace{-5pt}
\end{figure}

To address this problem, we introduce a \texttt{-legalize} \texttt{-dataflow} pass in ScaleHLS to analyze the dependencies between dataflow nodes and automatically legalize the targeted function. Fig. \ref{fig:dataflow}(a) shows an example dataflow containing five procedures, where each edge corresponds to a tensor delivering. We can observe that Fig. \ref{fig:dataflow}(a) is illegal as there is a path between \emph{Proc0} and \emph{Proc3} bypassing \emph{Proc1-2}. This dataflow can be conservatively legalized to Fig. \ref{fig:dataflow}(b) through the \texttt{-legalize-dataflow} pass. To eliminate the bypass path, \emph{Proc1-3} are organized into the same dataflow stage, thereby \emph{Proc0}, \emph{Proc1-3}, and \emph{Proc4} can construct a 3-stages dataflow. Note that in Fig. \ref{fig:dataflow}(b), the output buffers of \emph{Proc0} and \emph{Proc3} are automatically double buffered after the directive is successfully applied, with the cost of utilizing more memory resources than the original dataflow in Fig. \ref{fig:dataflow}(a).

Alternatively, the dataflow can be aggressively legalized to Fig. \ref{fig:dataflow}(c) through inserting \emph{Copy} nodes. The original bypass path is broken by the two inserted copy nodes, which enable a more fine-grained 5-stage dataflow. Assuming each procedure in the dataflow has a latency of $1t$, the conservative and aggressive legalization improves the dataflow interval from $5t$ to $3t$ and $1t$, respectively. However, the downside of the aggressive legalization is more computation and memory resources are consumed. The strategy of inserting copy nodes can be tuned through a \texttt{insert-copy} pass option. If the \texttt{insert-copy} option is enabled, copy nodes are inserted until the main path and the bypass path have the same number of nodes on them. Note that if the target function cannot be legalized, ScaleHLS will provide such diagnostics back to users and avoid unpredictable design transforms to be applied.

\subsubsection{Split Function}
Once the dataflow is legalized, the original function can be splitted into a top function and multiple sub-functions by a \texttt{-split-function} pass. Procedures or inserted copy nodes organized into the same dataflow stage can be safely clustered into a new sub-function and converted to a function call. At this stage, we find a throughput-area tradeoff space can be explored by merging adjacent dataflow stages into one. For example, in Fig. \ref{fig:dataflow}(d), every two adjacent stages are merged together, constructing a new 3-stages dataflow with less resource utilization compared to Fig. \ref{fig:dataflow}(c) and an interval of $2t$. To enable this design space, we define \emph{granularity} as the number of adjacent dataflow stages to be merged. The \texttt{-split-function} pass supports a \texttt{min-gran} parameter to specify the minimum granularity during the splitting. Therefore, at least \texttt{min-gran} adjacent dataflow stages are splitted into the one sub-function and converted to one function call.

\begin{figure*}
    \centering
    \includegraphics[width=\textwidth]{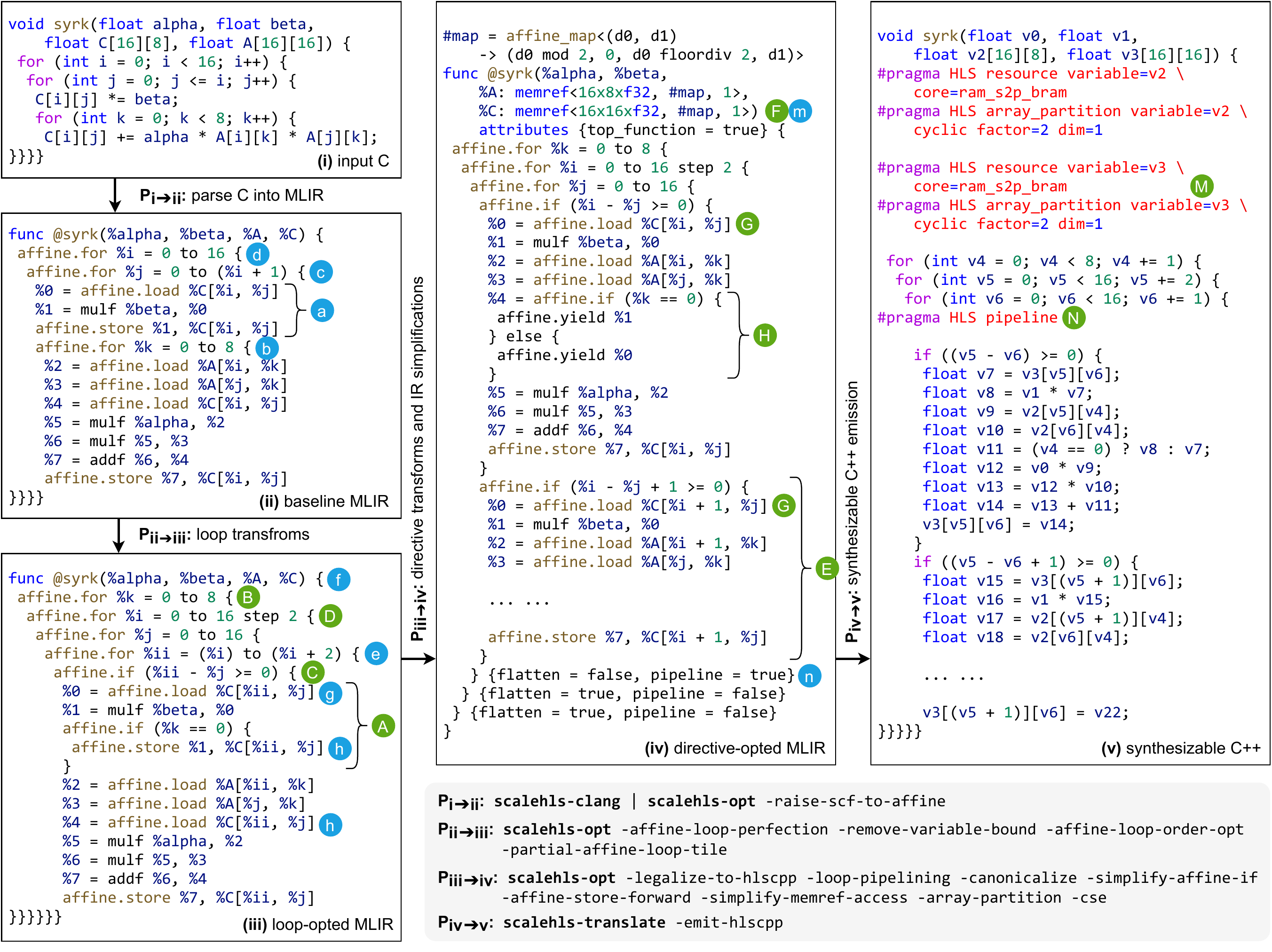}
    \vspace{-10pt}
    \caption{An SYRK computation kernel example. \texttt{scalehls-clang} compiles C program into the MLIR framework. \texttt{scalehls-opt} is the command line tool for conducting all conversion, transform, and analysis passes of ScaleHLS, while \texttt{scalehls-translate} is for the MLIR to C/C++ translation. Some operation attributes or types are omitted for simplicity.}
    \label{fig:opt}
    \vspace{-5pt}
\end{figure*}

\subsection{Loop Transform Passes}
\label{subsec:loop_transform}
We use the SYRK computation kernel shown in Fig. \ref{fig:opt} as the example in the following discussion. In this section, we introduce the loop transform passes provided by ScaleHLS, which corresponds to the $P_{ii \rightarrow iii}$ transformation of Fig. \ref{fig:opt}.

\subsubsection{Loop Perfectization}
Operations between loop statements, such as Fig. \ref{fig:opt}\textcircled{a} (hereinafter referred to as \ref{fig:opt}\textcircled{a}, \ref{fig:opt}\textcircled{b}, etc.), result in imperfect loops that may interfere with some important optimizations (e.g., loop tiling) and prevent the outer loops from being flattened for reducing latency. The \texttt{-affine-loop-perfectization} pass relocates the three in-between operations (\ref{fig:opt}\textcircled{a}) into the innermost loop and transforms them to \ref{fig:opt}\textcircled{\small{A}}, where all in-between operations are moved into a newly created \texttt{affine.if}. Then, operations except the state-modifying operations, such as stores, are hoisted out of the conditional.

\subsubsection{Loop Order Optimization}
Loop permutation can change the distance of loop-carried memory dependencies, thereby reducing the achievable $II$ of loop pipelining and reducing latency. The \texttt{-affine-loop-order-opt} pass can automatically perform affine-based memory dependency analysis and apply the best legal loop order to the targeted loop band. Specifically, loops associated with loop-carried dependencies are permuted to the outside in order to increase the distance of the dependencies. In the SYRK example, the original innermost \texttt{\%k}-loop (\ref{fig:opt}\textcircled{b}) is permuted to the outermost location (\ref{fig:opt}\textcircled{\small{B}}) by the loop order optimization pass. This pass also accepts an optional integers list, \texttt{perm-map}, allowing the loop order to be explicitly specified. The $i$-th element of \texttt{perm-map} indicates the new position of the $i$-th loop in the loop band, where positions are from the outermost loop to inner loops.

\subsubsection{Remove Variable Loop Bound}
Because MLIR focuses on rectangular iteration spaces, there are limitations on analyzing non-rectangular nested loops in MLIR. As a result, variable loop bounds may obstruct some loop optimizations and disrupt QoR estimation. The \texttt{remove-variable-bound} pass can calculate the minimum or maximum value of the expression of a variable loop bound as long as each item is a loop induction variable and has known lower and upper bounds. In the SYRK example, the variable loop bound of the \texttt{\%j}-loop (\ref{fig:opt}\textcircled{c}) is substituted with the constants and an \texttt{affine.if} operation (\ref{fig:opt}\textcircled{\small{C}}) is generated in the innermost loop for the conditional execution of the loop body. Although this pass may increase the overall iteration number of the loop band, it opens opportunities for subsequent optimizations which may offset the negative side effect.

\subsubsection{Loop Tiling}
Loop tiling is a common loop transform to improve data locality and accommodate the limited capacity of on-chip buffers. In the SYRK example, the \texttt{\%i}-loop (\ref{fig:opt}\textcircled{d}) is tiled with a factor of 2 and transformed into \ref{fig:opt}\textcircled{\small{D}}, and the generated intra-tile \texttt{\%ii}-loop is relocated into the innermost loop. The legality of loop tiling is validated through affine analysis before the transform is applied. The tiling size is determined by a \texttt{tile-size} parameter which can be tuned by the DSE engine.

\subsection{Directive Transform Passes}
\label{subsec:directive_transform}
In this section, we introduce the directive-level transform passes of ScaleHLS, which manipulate HLS-specific directives to further improve the design quality. The effect of the discussed passes are showcased in the $P_{iii \rightarrow iv}$ transformation of Fig. \ref{fig:opt}.

\subsubsection{Function and Loop Pipelining}
A legal pipeline directive allows no hierarchy in the target function or loop, thus all the sub-loops must be fully unrolled and all the sub-functions should be also pipelined~\cite{hls2020userguide}. The \texttt{-loop-pipelining} pass first attempts to legalize the targeted loop by fully unrolling all contained loops and pipelining all sub-functions. If the legalization succeeds, loop pipeline directive is applied to the target loop with the specified $II$. In the SYRK example, loop pipelining is applied to the \texttt{\%j}-loop and thus the contained \texttt{\%ii}-loop (\ref{fig:opt}\textcircled{e}) is fully unrolled and the duplicated loop body after loop unrolling is shown in \ref{fig:opt}\textcircled{\small{E}}. The \texttt{\%j}-loop is annotated as \texttt{pipeline} and all outer perfectly nested loops, \texttt{\%k} and \texttt{\%i}-loop, are annotated as \texttt{flatten}.

The \texttt{-function-pipelining} pass uses the same mechanism to legalize the targeted function before setting the function pipeline directive. Both the loop and function pipelining will recognize and diagnose illegal transform targets to avoid unpredictable compilation and allow specifying the targeted $II$ for exploring the tradeoff design space between throughput and resource utilization.

\subsubsection{Array Partition}
\label{subsec:array_partition}
ScaleHLS enhances the method proposed in~\cite{zhao2017comba} to automatically detect the memory access pattern of a program and apply the suitable array partition factor and fashion to each dimension of each array. The array partition metric $P$ of the $d$-th dimension of the $i$-th array can be represented with:

\begin{small}
\begin{equation}
    P_{i,d} = \frac{Accesses_i}{\underset{m,n}{max}(index^m_{i,d} - index^n_{i,d} + 1)},
\end{equation}
\end{small}
where $Accesses_i$ is the number of unique memory accesses in the targeted MLIR blocks, $index^m_{i,d}$ and $index^n_{i,d}$ are the indices of the $m$-th and $n$-th memory access operations. Note that $m$ and $n$ can be any two different memory accesses. The \texttt{-array-partition} pass applies \emph{cyclic} and \emph{block} partitions to the $d$-th dimension of the $i$-th array when $P_{i,d} >= 1$ and $P_{i,d} < 1$, respectively, with the partition factor set to $Accesses_i$. Taking the first dimension of the \texttt{\%C}-array (\ref{fig:opt}\textcircled{\small{F}}) as example, the index distance between the only two memory accesses (\ref{fig:opt}\textcircled{\small{G}}) is $(\%i + 1) - \%i + 1 = 2$. Therefore, we have $P=1$ and the applied partition fashion is \emph{cyclic}, which is encoded into the affine map of \texttt{\%C}-array.

As instantiated arrays can be accessed by sub-functions, an inter-procedural analysis is conducted to ensure: (1) the array partition directives are applied in the correct function scopes; (2) the globally optimal partition strategies are selected. The array partitioning process can also be guided by specifying the partition factors of each array which appears in the function through the \texttt{part-factors} parameter.

\subsection{IR Redundancy Elimination}
\label{subsec:simplification}
In addition to the graph, loop, and directive transforms, ScaleHLS adopts the methodology described in~\cite{aho1986compilers} and provides multiple passes to remove the redundant operations in the IR. The \texttt{-simplify-affine-if} pass eliminates dead branches of \texttt{affine.if} operations by detecting always-true/false conditions using affine analysis. The \texttt{-affine-store-forward} pass eliminates redundant memory access operations and unused memory instances through store-to-load forwarding. The \texttt{-simplify-memref-access} pass folds identical memory access operations if no dependency conflict exists. In the SYRK example, the memory access operations (\ref{fig:opt}\textcircled{h}) are eliminated and the IR is transformed to \ref{fig:opt}\textcircled{\small{H}}. ScaleHLS also exploits MLIR built-in passes, such as \texttt{-canonicalize} and \texttt{-cse} (common subexpression elimination)~\cite{mlirgithub}, to eliminate the redundancies in the IR and further optimize the quality of the HLS design.

\begin{figure}
    \centering
    \includegraphics[width=\linewidth]{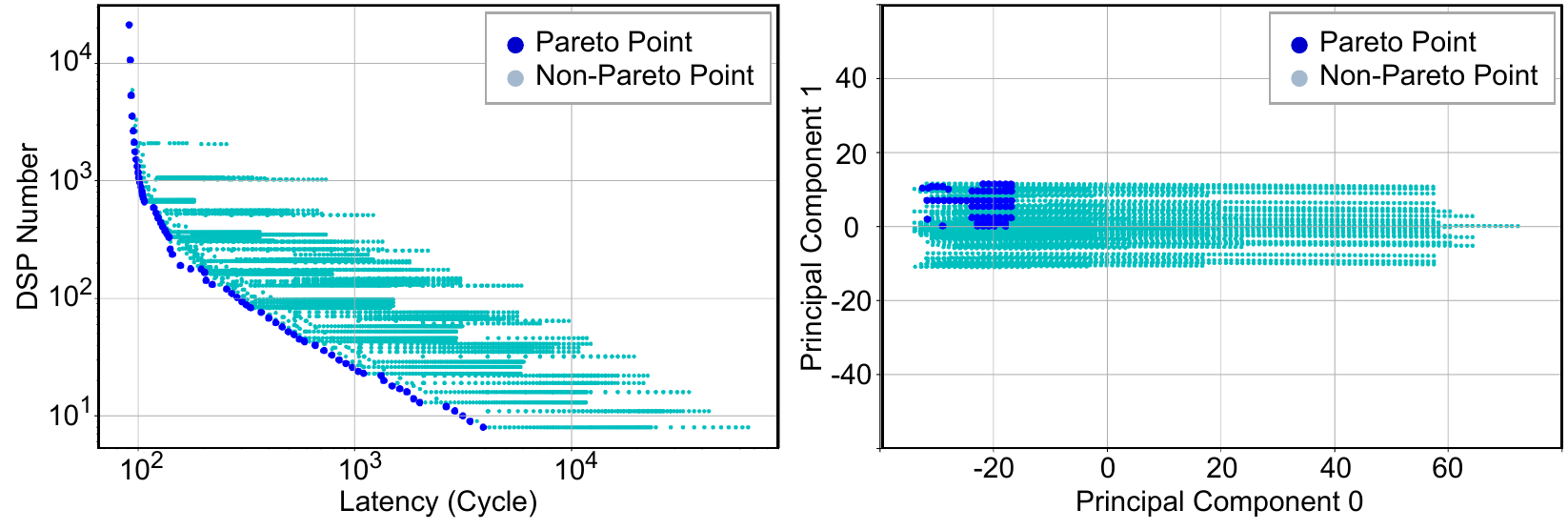}
    \vspace{-10pt}
    \caption{Design space profiling of a GEMM kernel. (a) the latency-area space; (b) PCA of the multi-dimensional design space.}
    \label{fig:dse_profile}
    \vspace{-5pt}
\end{figure}

\subsection{Automatic Design Space Exploration}
\label{subsec:dse}
On top of the representation and optimization of ScaleHLS, we can construct a multi-dimensional design space, where each dimension corresponds to the on/off or a tunable parameter of a transform pass.
% In addition, large HLS designs typically have complicated hierarchies - optimization targets (e.g., function or loop) nest with each other while every target is associated with a multi-dimensional design space. With regards to this, how to efficiently explore the hierarchical design space becomes the major challenge.
In this section, we propose an automated DSE engine assisted with an analytical model-based QoR estimator for exploring the design space. 

\subsubsection{QoR Estimation}
\label{subsec:qor_estimation}
The RTL generation downstream tools, such as Vivado HLS, can take minutes to hours to complete the compilation and to report the synthesis results, which (1) limits the total number of design points that can be evaluated during DSE, thus results in sub-optimal solutions and (2) significantly increases the DSE time to up to tens of hours. To solve these problems and rapidly evaluate the design points found by the DSE engine, we develop a QoR estimator to estimate the latency and resource utilization of the HLS designs. We adopt an ALAP (as late as possible) algorithm to schedule each MLIR block in the design. The memory ports are considered as non-shareable resources and constrained in the scheduling except between two or more memory read operations with identical address indices. The dependencies between operations are extracted through define-use and memory dependency analysis, where function calls and loops are viewed as nodes in the dependency graph. By accurately modeling the latency and resource utilization, ScaleHLS can estimate the effect of design transforms in the early stage of compilation, which makes the HLS optimizations more predictable.

% As discussed in Section \ref{subsec:loop_directive}, ScaleHLS directly unrolls the loops in the IR when loop unrolling directives are applied, thus loop unrolling does not need to be separately handled in the estimation. Since loop pipelining directives are represented with customized MLIR attributes, the estimator will parse the attribute and estimate the minimal $II$ if a loop is pipelined. We adopt the algorithm proposed in~\cite{rau1994iterative} for calculating the minimal $II$. We have:

% \vspace{-6pt}
% \begin{small}
% \begin{equation}
%     II_{min} = max(II^{res}_{min}, II^{dep}_{min}),
% \vspace{-3pt}
% \end{equation}
% \end{small}
% where $II^{res}_{min}$ and $II^{dep}_{min}$ are the minimal resource-constrained and dependency-constrained $II$, which can be calculated as:

% \vspace{-4pt}
% \begin{small}
% \begin{equation}
%     II^{res}_{min} = \underset{i,p}{max}\left( \left \lceil \frac{Accesses_{i,p}}{Ports_{i,p}} \right \rceil \right),
% \end{equation}
% \begin{equation}
%     \label{eq:dep_ii}
%     II^{dep}_{min} = \underset{d}{max}\left( \left \lceil \frac{Delay_d}{Distance_d} \right \rceil \right).
% \vspace{-5pt}
% \end{equation}
% \end{small}

% $Accesses_{i,p}$ and $Ports_{i,p}$ are the number of memory access operations and memory ports of the $p$-th partition of the $i$-th array. $Delay_d$ and $Distance_d$ are the scheduling delay and distance (calculated from the dependency vector) of each pair of loop-carried dependencies.

\subsubsection{DSE Algorithm}
\label{subsec:dse_algorithm}
The target of the DSE engine is to search for the Pareto frontier of the latency-area tradeoff space. By tuning the parameters of the transform passes shown in Tab. \ref{tab:passes}, we can construct a multi-dimensional design space for each input HLS design. Although the proposed QoR estimator can rapidly map a design point discovered in the multi-dimensional design space to the latency-area space, the powerful ScaleHLS transform passes can easily generate millions of design points, making exhaustive search impossible. The design space profiling of a GEMM (general matrix multiply) kernel~\cite{blackford2002updated} is shown in Fig. \ref{fig:dse_profile}, where we employ principle component analysis (PCA) for dimensionality reduction and exploratory analysis. We can observe that the Pareto points (deep blue) in the latency-area space are clustered in the PCA space. For instance, if pipeline $II=2$ is a Pareto point for a nested loop, there is a high possibility that its neighbors (e.g., pipeline $II=3$) are also Pareto points with different latency-area tradeoffs. Based on this observation, we design a 5-step neighbor-traversing algorithm to explore the design space.

In \textbf{Step1 Initial Sampling}, we randomly sample the whole design space and evaluate the latency and resource utilization of each sampled design point using the QoR estimator. Then, the initial Pareto frontier is extracted from all sampled points. In \textbf{Step2 Point Proposal}, we randomly select a design point in the current Pareto frontier and propose its closest neighbor as the new point to evaluate. This point proposal method can effectively inherit beneficial optimization parameters from evaluated design points. In \textbf{Step3 Point Evaluation}, we call the QoR estimator to evaluate the new design point proposed by Step2. The current Pareto frontier is then updated if any point in the current frontier is dominated by the new design point. In \textbf{Step4 Frontier Evolution}, we repeat Step2 and Step3 until no eligible neighbor can be found or meeting the early-termination criteria (e.g., maximum iteration number). In this procedure, the discovered Pareto frontier evolves in each iteration and progressively approaches the "real" Pareto frontier. In \textbf{Step5 Design Finalization}, we sort the discovered Pareto points in ascending order of latency and then select the first point meeting the resource constraints as the final solution. The finalized design is emitted as synthesizable C++ code. This DSE algorithm is implemented as an MLIR transform pass called \texttt{-dse} which can be applied on the input HLS designs without any manual efforts. The evaluation of the DSE engine is performed in Section \ref{sec:results}. Note that given the HLS transform and analysis library of ScaleHLS, the DSE engine is extensible to support different optimization algorithms.

\section{End-to-End Integration}
\label{sec:integration}

\subsection{HLS C Front-end}
\label{subsec:front_end}
The C front-end takes synthesizable HLS C code and emits the corresponding MLIR in the \texttt{scf} dialect. The \texttt{scf} dialect provides an abstraction for static control flow and has a similar set of operations to statements in C, which reduces the analysis process in the front-end. For instance, a \texttt{for} loop can be directly translated to an \texttt{scf.for} operation. The output in the \texttt{scf} dialect is then \emph{raised} into the \texttt{affine} dialect using an ScaleHLS pass called \texttt{-raise-scf-to-affine}. This pass checks whether an \texttt{scf.for} operation is an affine loop and translates it into an \texttt{affine.for} operation if it is. Otherwise, the loop remains as an \texttt{scf.for} operation. Also, the MLIR pass raises each memory statement to an \texttt{affine} operation if its address indices have affine formats. The $P_{i \rightarrow ii}$ transformation of Fig. \ref{fig:opt} shows the procedure of parsing HLS C codes into the MLIR framework.

MLIR has its unique memory and indexing types. First, the memory type \texttt{memref} is a set of exclusive pointers to the memory and size parameters of the memory~\cite{lattner2020mlir}. The \texttt{memref} type solves delinearization problem of parametrically sized arrays, which was not well-supported in LLVM~\cite{grosser2015optimistic}. The translation to \texttt{memref} is simplified in our front-end because common HLS tools, such as Vivado HLS, only accepts a subset of C~\cite{hls2020userguide}. For instance, all the arrays have to have fixed sizes. These types are directly translated to fixed-size \texttt{memref} types. A pointer that points to a scalar has a 1 $\times$ 1 \texttt{memref} type in MLIR. If an unsupported struct such as pointer to pointer is found, the input code is rejected by the C front-end. 

% Second, the \texttt{index} type in MLIR is an integer type with a platform-specific bit width. The \texttt{index} types are typically used for a set of constructs such as loop iterators and memory indices. In the translation, our C front-end checks whether an integer variable can be used as one of these constructs. For instance, the iterator in \texttt{index} type of a \texttt{for} loop should not overflow, otherwise the loop is translated into an \texttt{scf.while}.

\subsection{HLS C/C++ Code Emission}
\label{subsec:emitter}
After the completion of all conversions and optimizations, the structured IR can be emitted as synthesizable C/C++ code for generating the RTL code. The $P_{iv \rightarrow v}$ transformation of Fig. \ref{fig:opt} shows the MLIR to C++ emission of the SYRK example. The HLS C/C++ emitter of ScaleHLS requires the control flow to be represented by \texttt{affine} or \texttt{scf} operations. Then, it can directly translate \texttt{affine/scf.for} and \texttt{if} operations to the \texttt{for} and \texttt{if} statements in C/C++. The array partition, resource, and interface information is decoded from the type of memories (\ref{fig:opt}\textcircled{m}) and emitted as \texttt{pragma} directives (\ref{fig:opt}\textcircled{\small{M}}). Meanwhile, the applied HLS-specific optimizations represented as attributes (\ref{fig:opt}\textcircled{n}) are also parsed by the emitter accordingly and inserted into the corresponding code region. Notably, to ensure the synthesizability of the generated C/C++, the emitter always converts returned values to input pointers.

\begin{table*}
    \centering
    \caption{DSE results of large-scale computation kernels.}
    \vspace{-5pt}
    \label{tab:kernel_results}
    \begin{tabular}{c|c|c|c|c|c|c|c|c}
        \midrule
        \textbf{Kernel} & \textbf{Prob. Size} & \textbf{Speedup} & \textbf{LP} & \textbf{RVB} & \textbf{Perm. Map} & \textbf{Tiling Sizes} & \textbf{Pipeline II} & \textbf{Array Partition Factors} \\
        \midrule
        \textbf{BICG} & 4096 & 41.7$\times$ & No & No & [1, 0] & [16, 8] & 43 & $A$:[8, 16], $s$:[16], $q$:[8], $p$:[16], $r$:[8] \\
        % \hline
        \textbf{GEMM} & 4096 & 768.1$\times$ & Yes & No & [1, 2, 0] & [8, 1, 16] & 3 & $C$:[1, 16], $A$:[1, 8], $B$:[8, 16] \\
        % \hline
        \textbf{GESUMMV} & 4096 & 199.1$\times$ & Yes & No & [1, 0] & [8, 16] & 9 & $A$:[16, 8], $B$:[16, 8], $tmp$:[16], $x$:[8], $y$:[16] \\
        % \hline
        \textbf{SYR2K} & 4096 & 384.0$\times$ & Yes & Yes & [1, 2, 0] & [8, 4, 4] & 8 & $C$:[4, 4], $A$:[4, 8], $B$:[4, 8] \\
        % \hline
        \textbf{SYRK} & 4096 & 384.1$\times$ & Yes & Yes & [1, 2, 0] & [64, 1, 1] & 3 & $C$:[1, 1], $A$:[1, 64] \\
        % \hline
        \textbf{TRMM} & 4096 & 590.9$\times$ & Yes & Yes & [1, 2, 0] & [4, 4, 32] & 13 & $A$:[4, 4], $B$:[4, 32] \\
        \midrule
        \multicolumn{9}{l}{\tabincell{l}{The data types of all kernels are 32-bits floating-points. \emph{Speedup} is with respect to the baseline designs from PolyBench-C without the optimization of \\ DSE. \emph{LP} and \emph{RVB} denote \emph{Loop Perfectization} and \emph{Remove Variable Bound}, respectively. In the \emph{Loop Order Optimization}, the $i$-th loop in the loop nest \\ is permuted to location $PermMap[i]$, where locations are from the outermost loop to inner.}}
    \end{tabular}
\end{table*}

\begin{figure*}
    \centering
    \includegraphics[width=\textwidth]{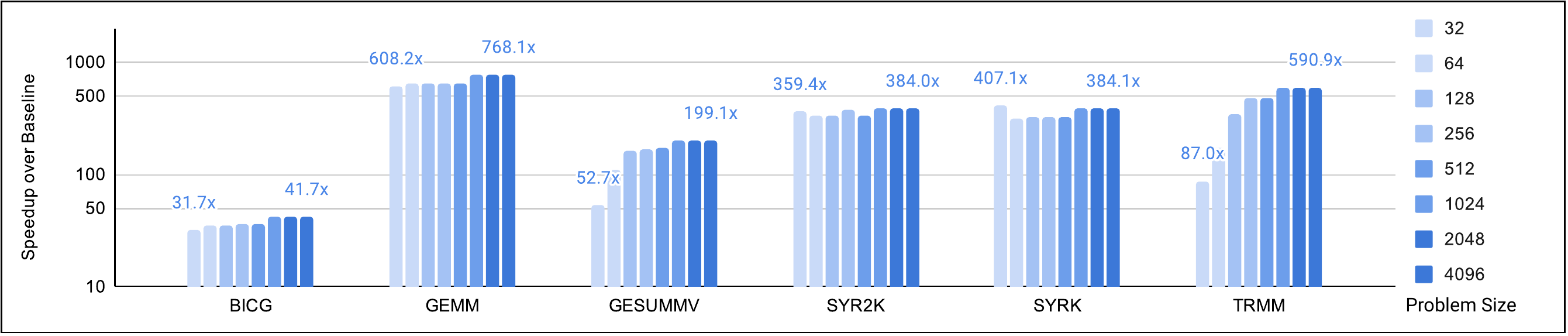}
    \vspace{-10pt}
    \caption{Scalability study of computation kernels. The problem sizes of computation kernels are scaled from 32 to 4096 and the DSE engine is launched to search for the optimized solutions under each problem size.}
    \label{fig:kernel_results}
    \vspace{-5pt}
\end{figure*}

\section{Experimental Results}
\label{sec:results}
To evaluate the ScaleHLS compilation framework, we conduct comprehensive experiments and ablation studies in this section. Xilinx Vivado HLS 2019.1 is adopted for generating RTL code. All reported performances and resources utilization are collected from the synthesis results reported by Vivado HLS.

\subsection{Large-Scale Computation Kernels}

\subsubsection{Automatic DSE results}
We evaluate the DSE engine on six different computation kernels (BICG, GEMM, GESUMMV, SYR2K, SYRK, and TRMM) picked from PolyBench-C \cite{pouchet2012polybench} with a problem size of 4096. The target platform is Xilinx XC7Z020 FPGA, which is an edge FPGA with 4.9 Mb memories, 220 DSPs, and 53,200 LUTs. The resource constraints and non-optimized computation kernels written in C are passed into the DSE engine, which is then launched to search for the optimized solutions. Finally, the generated designs are evaluated and the results are shown in Tab. \ref{tab:kernel_results}. Among all six benchmarks, a speedup ranging from 41.7$\times$ to 768.1$\times$ is obtained compared to the baseline design, which is the original computation kernel from PolyBench-C without the optimization of DSE or manual code rewriting. Tab. \ref{tab:kernel_results} also lists the parameters selected by DSE for each transform pass. Notably, in the procedure of loop tiling, all generated intra-loops are absorbed into the innermost loop region and fully unrolled for increasing the computation parallelism.

After studying the solutions discovered by the DSE engine, we find the performance gains come from multiple sources: (1) loop perfectization and variable loop bound removal regularize the target loop bands and enable the subsequent optimizations; (2) loop permutation alleviates (or eliminates) the impact of memory dependencies and improves the achievable pipeline $II$ by reducing the distance of loop-carried dependencies; (3) the computation parallelism and resource utilization are increased through loop tiling and intra-tile loop unrolling; (4) loop pipelining is applied and the target $II$ is fine-tuned to tradeoff between resource-sharing and throughput while accommodating the resource constraints; (5) array partitioning strategies are automatically selected to match the memory access patterns after loop transformations. The BICG benchmark cannot benefit from loop permutation because every loop in the loop nests is associated with critical loop-carried dependency which prevents the DSE engine to effectively reduce the pipeline $II$. However, the DSE engine still discovers a reasonable solution for the BICG benchmark and achieves a 41.7$\times$ speedup through increasing the computation parallelism. Benchmarks except BICG benefit from all speedup sources above, and achieve significant performance improvements under the constrained on-chip resources available on the edge FPGA platform.

\begin{table}
    \caption{Case study of GEMM kernel with a problem size of 4096.}
    \label{tab:gemm_study}
    \vspace{-5pt}
    \centering
    \begin{tabular}{c|c|c|c}
        \midrule
        \textbf{Design} & \textbf{Cycles} & \textbf{Speedup} & \textbf{DSP (Util. \%)} \\
        \midrule
        \textbf{Unoptimized} & $1.237\times10^{12}$ & $1.0\times$ & $5~(2.3\%)$ \\
        \textbf{DSE Optimized} & $1.610\times10^9$ & $768.1\times$ & $217~(98.6\%)$ \\
        \textbf{Manually Optimized} & $2.684\times10^9$ & $460.9\times$ & $220~(100.0\%)$ \\
        \textbf{Theoretical Bound} & $1.562\times10^9$ & $791.9\times$ & $220~(100.0\%)$ \\
        \midrule
    \end{tabular}
    \vspace{-5pt}
\end{table}

To better understand the quality of solutions found by the DSE, we perform a case study on the GEMM kernel. We manually implement an HLS design on the same FPGA platform using a rich set of directives as well as code-rewriting driven by human design experience and expertise. We also calculate the best achievable latency on the targeted FPGA by assuming all DSPs on chip run without any stall and the kernel can be perfectly parallelized. The case study results are shown in Tab. \ref{tab:gemm_study}. We can observe that the HLS design generated by our DSE achieves 0.97$\times$ of the theoretical bound and is around 1.67$\times$ better than the manually optimized HLS design. Note that the DSE only takes minutes to find the design, while the manual design takes us about 10 hours to finish.

\subsubsection{Comparison with Previous Works}
Previous efforts~\cite{zuo2015polyhedral, zhong2016lin, zhao2017comba} also investigate automatic DSE methods to optimize computation kernel level algorithms. However, they only support directive optimizations, thus are difficult to comprehensively explore the design space and find reasonable design points when the problem sizes are large. For example, on the six scaled-up benchmarks shown in Tab. \ref{tab:kernel_results}, the open-sourced framework~\cite{zhao2017comba} either generates solutions that cannot be synthesized by Vivado HLS or takes an unreasonable long time on exploring the large design spaces. Meanwhile, as previous DSE efforts do not support the transform and analysis library featured by ScaleHLS, they still rely on human to provide optimization hints or rewrite the code before launching the DSE, leading to low-efficient and partially-automated compilation flows. Our multi-level representation and automated optimization enable ScaleHLS to find previously unachievable design points, explore a more comprehensive design space, and directly generate synthesizable HLS designs.

\begin{table*}
    \centering
    \caption{Optimization results of representative DNN models.}
    \vspace{-5pt}
    \begin{tabular}{c|c|c|c|c|c|c|c}
        \midrule
        \textbf{Model} & \textbf{Speedup} & \tabincell{c}{\textbf{Runtime} \\ \textbf{(seconds)}} & \tabincell{c}{\textbf{Memory} \\ \textbf{(SLR Util. \%)}} & \tabincell{c}{\textbf{DSP} \\ \textbf{(SLR Util. \%)}} & \tabincell{c}{\textbf{LUT} \\ \textbf{(SLR Util. \%)}} & \tabincell{c}{\textbf{Our DSP Effi.} \\ \textbf{(OP/Cycle/DSP)}} & \tabincell{c}{\textbf{DSP Effi. of} \\ \textbf{TVM-VTA~\cite{moreau2018vta}}} \\
        \midrule
        \textbf{ResNet-18} & 3825.0$\times$ & 60.8 & 91.7Mb (79.5\%) & 1326 (58.2\%) & 157902 (40.1\%) & 1.343 & 0.344 \\
        \textbf{VGG-16} & 1505.3$\times$ & 37.3 & 46.7Mb (40.5\%) & 878 (38.5\%) & 88108 (22.4\%) & 0.744 & 0.296 \\
        \textbf{MobileNet} & 1509.0$\times$ & 38.1 & 79.4Mb (68.9\%) & 1774 (77.8\%) & 138060 (35.0\%) & 0.791 & 0.468 \\
        \midrule
        \multicolumn{8}{l}{\tabincell{l}{\emph{Speedup} is with respect to the baseline designs compiled from PyTorch by ScaleHLS but without the multi-level optimization.}}
    \end{tabular}
    \label{tab:dnn_results}
\end{table*}

\begin{figure*}
    \centering
    \includegraphics[width=\textwidth]{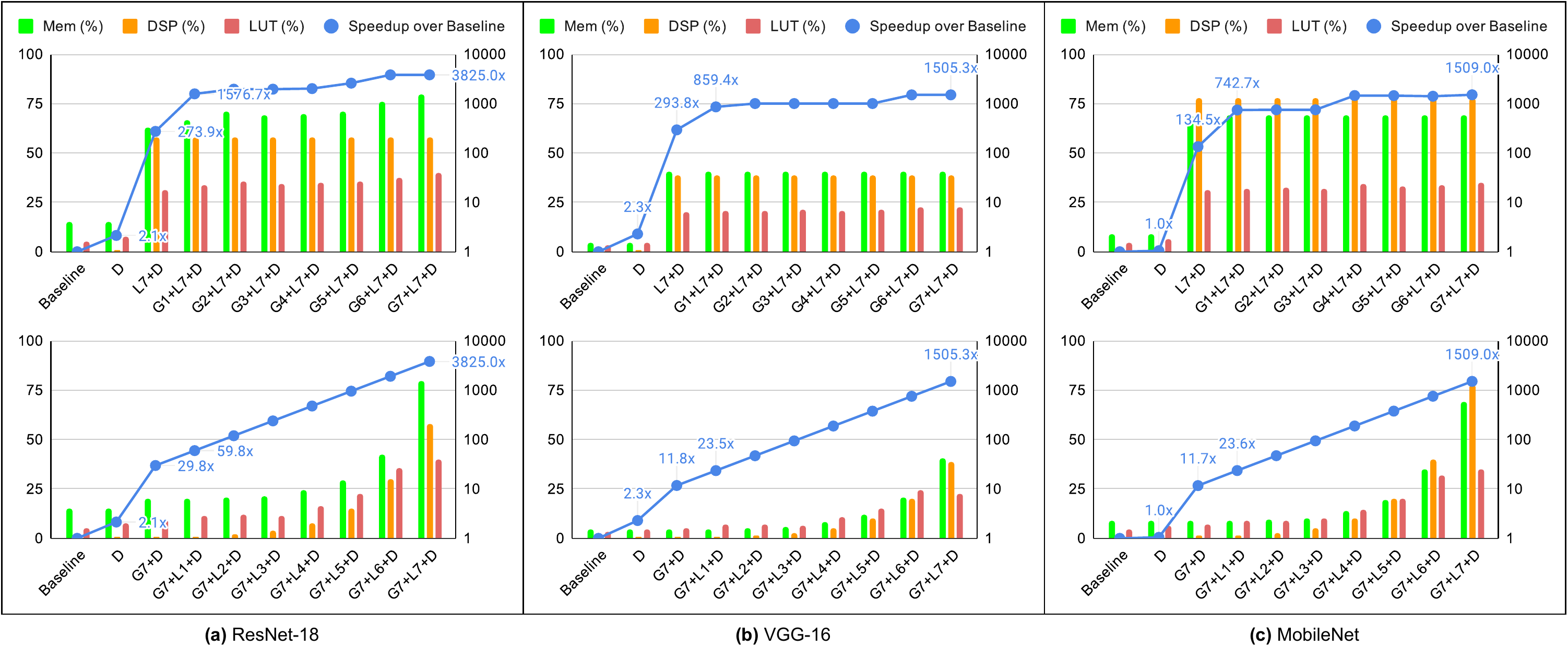}
    \vspace{-10pt}
    \caption{Ablation study of DNN models. $D$, $L\{n\}$, and $G\{n\}$ denote directive, loop, and graph optimizations, respectively. Larger $n$ indicates larger loop unrolling factor and finer dataflow granularity for loop and graph optimizations, respectively.}
    \label{fig:dnn_results}
    \vspace{-5pt}
\end{figure*}

\subsubsection{Scalability Study}
To understand the performance of our framework on different problem sizes, we scale the problem sizes of the six benchmarks from 32 to 4096 and launch the DSE engine to search for the optimized solution under each setting. Fig. \ref{fig:kernel_results} shows the experimental results. We can observe that for BICG, GEMM, SYR2K, and SYRK benchmarks, the DSE engine can achieve stable speedup under all problem sizes. For GESUMMV and TRMM, the speedups for small problem sizes are lower because the small design space prevents the DSE engine from fully utilizing the available on-chip resources. Overall, our framework shows a strong scalability and can effectively optimize computation kernel level algorithms on a wide range of problem sizes.

\subsection{Large and Complicated Algorithms}

\subsubsection{Optimization Results}
We experiment the ability of handling large and complicated HLS designs of ScaleHLS by evaluating three representative DNN (deep neural networks) models for the CIFAR-10~\cite{krizhevsky2009learning} image classification task, ResNet-18~\cite{he2016deep}, VGG-16\cite{simonyan2014very}, and MobileNet~\cite{howard2017mobilenets}. These DNN models are constructed with a large number of different hidden layers and have sophisticated inter-layer dependencies. The target platform is one SLR (super logic region) of Xilinx VU9P FPGA which is a large FPGA containing 115.3 Mb memories, 2280 DSPs and 394,080 LUTs on each SLR. The PyTorch~\cite{paszke2019pytorch} implementations are parsed into ScaleHLS and optimized using the proposed multi-level optimization methodology. Graph, loop, and directive optimization passes are applied sequentially to improve the design quality at the corresponding IR level. The experimental results are shown in Tab. \ref{tab:dnn_results}. We can observe that by combining three levels of optimization, the generated HLS designs achieve significant speedups ranging from 1505.3$\times$ to 3825.0$\times$ on the metric of throughput compared to the baseline designs, which are compiled from PyTorch to HLS C/C++ through ScaleHLS but without the multi-level optimization applied. Notably, as shown in Tab. \ref{tab:dnn_results}, ScaleHLS only consumes 37.3 to 60.8 seconds to optimize the large HLS designs with a single line of command, which demonstrates the efficiency and scalability of our optimization methodology. The runtime is collected by using \texttt{-pass-timing}, a built-in statistic pass provided by the MLIR framework.

\subsubsection{Comparison with Previous Works} To the best of our knowledge, ScaleHLS is the first general-purpose HLS flow which can optimize and generate ResNet-18 level DNN accelerators without human-designed IPs or templates. Previous HLS optimization flows~\cite{zuo2015polyhedral, zhong2016lin, zhao2017comba} focus on small-scale algorithms, while compilation flows dedicated for DNNs rely on pre-defined IP libraries~\cite{umuroglu2017finn, zhang2018dnnbuilder, moreau2018vta} or parameterized templates~\cite{ye2020hybriddnn, meng2021dynamap} to generate the accelerator, which can not be generalized to applications other than DNNs. To better understand the optimization results of DNN models, we compare the DSP efficiency with TVM-VTA~\cite{moreau2018vta}, a widely accepted DNN acceleration framework written in HLS. DSP efficiency is a common metric for comparing the efficiency of DNN accelerators across different platforms, which can be calculated as:

\begin{small}
\begin{equation}
    Effi_{DSP}=\frac{OP/s}{Num_{DSP}\times Freq}
\end{equation}
\end{small}

As shown in Tab. \ref{tab:dnn_results}, ScaleHLS reaches a better DSP efficiency, while saves hundreds of human hours for designing the dedicated hardware IPs. These experimental results demonstrate that ScaleHLS can achieve fruitful productivity improvement on large and complicated algorithms.

\subsubsection{Ablation Study}
To quantify the speedup contributed by each of the three optimizations (directive, loop, and graph), we perform an ablation study shown in Fig. \ref{fig:dnn_results}. We can observe that the directive ($D$), loop ($L7$), and graph ($G7$) optimizations contribute 1.8$\times$, 130.9$\times$, and 10.3$\times$ average speedups on the three DNN benchmarks, respectively, demonstrating the effectiveness of our multi-level optimization methodology. Note that because the effect of array partitioning is larger as the loop unrolling factors increase, the actual speedup of directive optimizations are larger than 1.8$\times$ when combining with loop optimizations. ScaleHLS allows to tune the optimization level $n$ between 1 to 7 for loop and graph optimizations, which enables to explore the tradeoff space between area and speedup. Larger $n$ indicates larger loop unrolling factor and finer dataflow granularity for loop and graph optimizations, respectively, leading to higher throughput and more on-chip resources utilization. By comparing the speedup achieved by $G1+L7+D$ and $G7+L7+D$, we can observe that the speedup margin between $G1$ and $G7$ is 2.1$\times$ on average. Similarly, the speedup margin between $L1$ and $L7$ is 64.0$\times$ on average.

\section{Conclusion}
\label{sec:conclusion}
This paper presents ScaleHLS, an MLIR-based HLS compilation flow, which features multi-level representation and optimization of HLS designs and supports a transform and analysis library dedicated for HLS. ScaleHLS enables an end-to-end compilation pipeline by providing an HLS C front-end and a C/C++ emission back-end. An automated and extensible DSE engine is developed to search for optimized solutions in the multi-dimensional design spaces. Experimental results demonstrate that ScaleHLS has strong scalability to optimize large-scale and sophisticated HLS designs and achieves significant performance and productivity improvements on a set of benchmarks. In addition, ScaleHLS is an open-source project and we hope ScaleHLS could become an advanced open infrastructure of new HLS research in the future and boost the innovation in this area to face new challenges.

ScaleHLS leaves several directions for future works: (1) IP integration. The graph-level IR of ScaleHLS opens the opportunity to integrate existing hardware IPs into the compilation flow, making the integration and optimization of IPs an interesting research direction. (2) DSE algorithms. The transform and analysis library provided by ScaleHLS enables a great opportunity to further investigate the optimization algorithms for the multi-dimensional DSE problem of HLS. (3) Machine-learning based QoR estimation. Machine-learning methods can potentially capture more features from the hierarchical IR of ScaleHLS, thereby generating better estimation results than the analytical model-based methods. (4) RTL code generation within MLIR. Currently ScaleHLS leverages external HLS tools for generating the RTL code. However, a direct RTL code generation within MLIR can keep more information from the higher level IR and exploit the RTL-level representation and optimization (CIRCT~\cite{circtgithub}) to further improve the quality of the accelerator designs.

\section*{Acknowledgements}
We thank Eric Cheng of Laboratory for Physical Sciences (LPS) and Samuel Bayliss of Xilinx for insightful discussions. This work is supported in part by Xilinx Center of Excellence at UIUC, Xilinx Adaptive Compute Cluster (XACC) initiative, and BAH HT 15-1158 contract.


\bibliographystyle{IEEEtran}
\bibliography{refs}

\end{document}